\documentclass[preprint,12pt]{elsarticle}

\journal{Computational Materials Science}
\usepackage{amssymb}
\usepackage{xcolor}
\usepackage{threeparttable}

\usepackage{hyperref}
\hypersetup{breaklinks=true}
\usepackage{xurl}

\begin{document}

\begin{frontmatter}

\title{Prediction of coherent interfaces between diamond and clathrate structures}

\author[1]{Eva Posp\' i\v silov\' a}
\affiliation[1]{organization={Institute of Physics, Slovak Academy of Sciences},
  addressline={Dubravska cesta 11},
  city={Bratislava},
  postcode={84511},
  country={Slovakia}
}
\author[1,2]{Marek Mihalkovi\v c}
\affiliation[2]{organization={Centre of excellence for Advanced Materials, Slovak Academy of Sciences},
  addressline={Dubravska cesta 11},
  city={Bratislava},
  postcode={84511},
  country={Slovakia}
}
\date{\today}

\begin{abstract}

Diamond (or its binary zincblende variant)--type structure can form
coherent interface with clathrate type II via the common
transitional layer known previously as a 3x3 dimer-stacking fault (DS)
reconstruction of the (111)-diamond surface. The generic $\sim$11\%
lattice misfit can be eliminated in multicomponent heterostructures
such as Ge(diamond)/CsSn(clathrate) or InN(zincblende)/Ge(clathrate).
Interface models subjected to ab--initio molecular dynamics
annealing are stable up to the temperatures approaching melting point
of the constituent systems, and in some studied cases the
diamond/clathrate bonding is stronger than the intra-clathrate
bonding, as evidenced by simulated crack experiments.
Composition-calibrated lattice--matching can stabilize even metastable
clathrates as epitaxially grown films on the diamond/zincblende
substrate.

\end{abstract}

\begin{graphicalabstract}
\includegraphics[width=5.5in,angle=0]{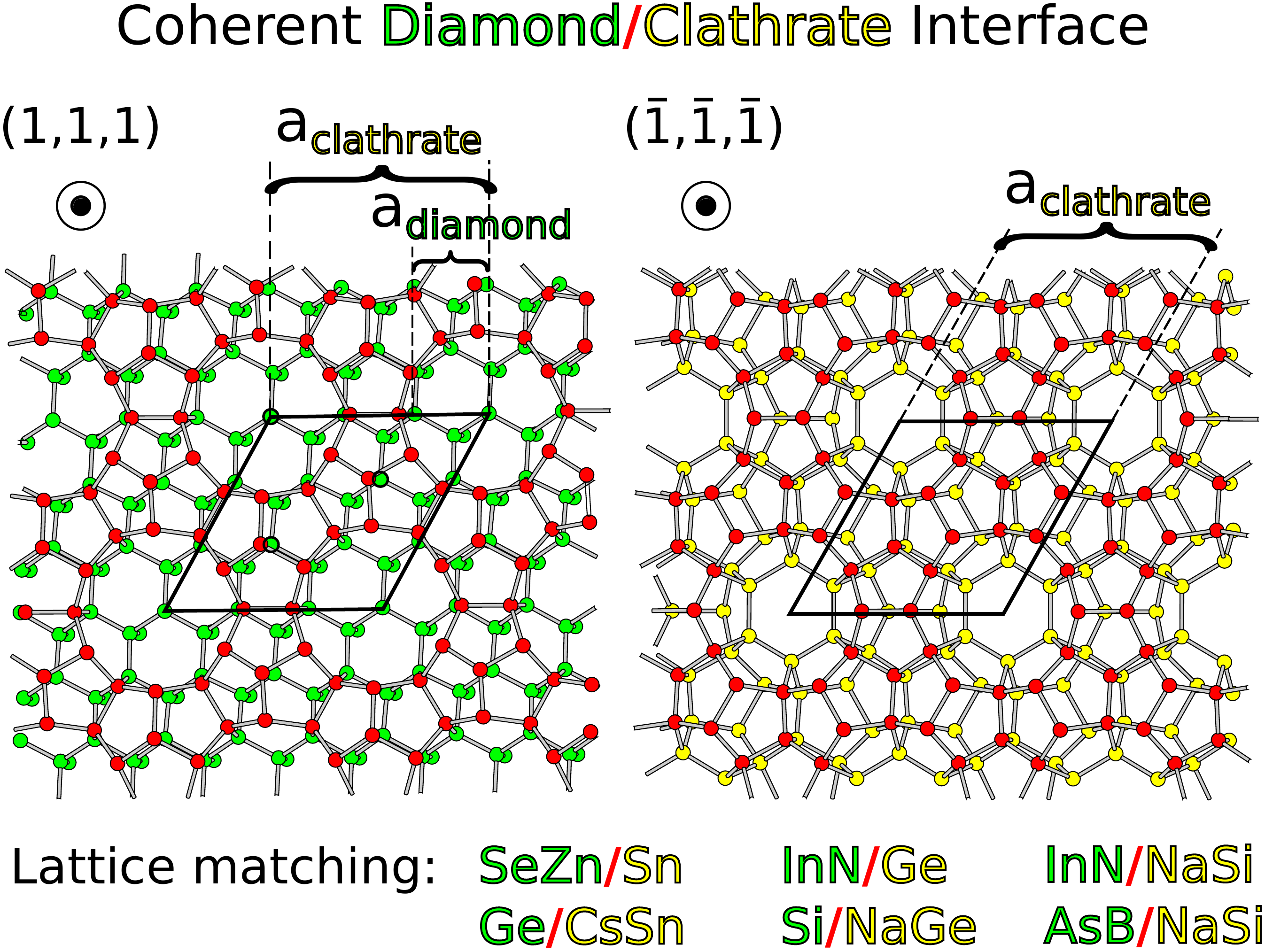}
\end{graphicalabstract}

\begin{highlights}
\item diamond (or zincblende) and clathrate structures can form effective coherent interface
\item design by chemical composition can stabilize metastable clathrate films on diamond substrate
\item lattice-matched diamond/clathrate or zincblende/clathrate interface composition examples: AsB/NaSi, InN/NaSi, InN/Ge, Si/NaGe, SeZn/Sn, Ge/CsSn
\item interface scrutinized by DFT : high-temperature molecular dynamics annealing, tensile stress testing up to the cracking
\end{highlights}


\end{frontmatter}

\section{Introduction}
\label{sec:intro}

  $sp^3$--bonded materials are the basis for semiconductor or
  optoelectronic applications in their diamond-structure or binary
  zincblende forms, while the inorganic clathrates stabilized by
  (large) guest atoms are a prominent thermoelectric material.
  Coherent coexistence of the two $sp^3$ systems would open an
  interesting property-tuning possibility\cite{martinez2013}, but
  their incompatible architecture (diamond atoms constitute planar
  sheets, clathrates are variable assemblies of approximately
  spherical cages) seemingly prevents it.

Nevertheless, the recent
literature contains three reports of a spontaneous formation of the
clathrate/diamond {\em coherent} interface, and robustness of the
coherency is manifested in diversity of the three situations:
intermetallic Ge(diamond)/NaGe(clathrate) prepared by forming
clathrate by high-temperature annealing of NaGe precursor phase on
(111)-diamond Ge surface~\cite{kume_gena}; second, by molecular
dynamics simulations leading to ice crystallization from the melt on
the surface of clathrate hydrate~\cite{iceclath}; or finally in
molecular dynamics simulations of colloidal tetrahedral patchy
particles system~\cite{patchypart}. None of these cases lead to
explicit understanding of the interface structure: in case of Ge/NaGe
experiment, authors hypothesized existence of a ``thin buffer layer''
between the diamond and clathrate; in case of clathrate hydrate
atomistic simulation, due to the lattice mismatch the surface
periodicity was dictated by stripes of disordered structure, while the
coherent patches were too small to indicate supercell size. Finally,
Ref.~\cite{patchypart} in their Fig.(2h) showed coherent interface
diamond/clathrate apparently in the same orientational relationship as
the previous case, but did not reveal or even analyze its
structure.

This paper clarifies surprising relationship between seemingly
incompatible diamond and clathrate structures: the so called
3$\times$3-$DS$ reconstruction of the diamond(111) surface layer
exactly coincides with a layer in type II or type III clathrate.  The
coincidence warrants perfect bond saturation for clathrate atoms, and
good (2/3 fraction) bond saturation on the diamond side of the
interface. The $\sim$11\% generic mismatch at the coincidence layer
can act as a lever providing lattice-matching advantage stabilizing
clathrate against the competing diamond structure, by selecting
appropriate compositions of the two interface components.


In the following, we systematically screen the space of 
$sp^3$--bonded chemical systems for compositions satisfying the
diamond/clathrate interface lattice-matching condition, and we
demonstrate robustness and stability of the selected interfaces by
subjecting them to DFT uniaxial stretching and high--temperature
molecular dynamics annealing.

\section{Methods}
\label{sec:methods}

In order to optimize and test models of the interface, we
employ {\em ab--initio} density--functional theory setup as
implemented in plane--wave package VASP~\cite{Kresse96}, using
projector augmented wave method potentials~\cite{PAW} and Perdew-Wang
density functional~\cite{PW91} ``PW91'' utilizing generalized gradient
approximation. As a default setup, we use VASP ``Accurate'' setting,
and converge energy on $k$-point mesh density to 1 meV/atom accuracy.

We also perform uniaxial tensile testing of the interfaces by applying
strain normal to the interface. We start from fully relaxed structure,
and iteratively increase tensile strain normal to the interface
plane. The straining step was 1 or 2\%, followed by fixed-volume
relaxation (``ISIF=2'' VASP setting). The next iteration starts
from preceding relaxed structure.

To diagnose possibly metastable local atomic configurations at the
interface, we anneal models in {\em ab--initio} molecular dynamics
(AIMD) simulation at gradually increasing temperature, until they
eventually melt. Since AIMD annealing is time consuming, we
  experimented to minimize the number of independent k-points required
  to maintain plausible accuracy. For our samples containing
  $\sim$100-200 atoms in hexagonal cell with 10-15\AA~ length of
  $a$-parameter and $c$-parameter at least 1.7-times longer, we
  converged on using just three manually set k-points:
  $(0.1,0.1,0.1)$, $(1/4,1/3,1/4)$ and $(-1/3,1/4,1/4)$ with equal
  weighting factors (no symmetry). Three selected systems were
  scrutinized for accuracy of this setting: Ge/CsSn, InN/NaSi and
  AsB/NaSi. In each case we selected randomly 5 snapshot samples
  separated by at least 2000 steps in high-temperature simulations,
  and compared energies with calculation using (3$\times$3$\times$2)
  Monkhorst mesh (10 independent k-points). We found systematic energy
  differences $\Delta E\sim 8.5\pm 0.4$ meV/atom for AsB/NaSi system,
  $3.3\pm 0.5$ meV/atom for InN/NaSi and $-0.5\pm 2.0$ meV/atom for
  Ge/CsSn. The quoted r.m.s. deviations suggest essentially no effect
  of the reduced k-point set on the accuracy; largest r.m.s. occured
  for system with Sn that has pronounced metallization tendency hence
  convergence on k-points may be slower. We also monitored state of a
sample by checking time evolution of the total energy (melting or
substantial structural change leads to abrupt change of total
potential energy), and by visual inspection of snapshot
samples. At highest temperatures (above 1500K) the InN/NaSi and
  AsB/NaSi simulation occasionally failed to converge electronic
  self-consistency cycle; setting the time step to 5fs and increasing
  energy cutoffs (ENCUT variable) from 319 to 350 eV (InN/NaSi), or
  from 400 to 500 eV (InN/NaSi) removed the problem. Typical
trajectories performed at a temperature were 50$ps$ (10000 ionic
steps). If the annealing at a temperature $T$ did not lead to
structural changes, we increased the temperature by 50K.  Once the
sample melted, we quenched it to obtain energy difference $\Delta E_a$
between the diamond/clathrate vs diamond/amorphous interfaces.

\section{Interface}
\label{sec:ifejs}

\begin{figure}[t]
  \centering
  \includegraphics[width=5in,angle=0]{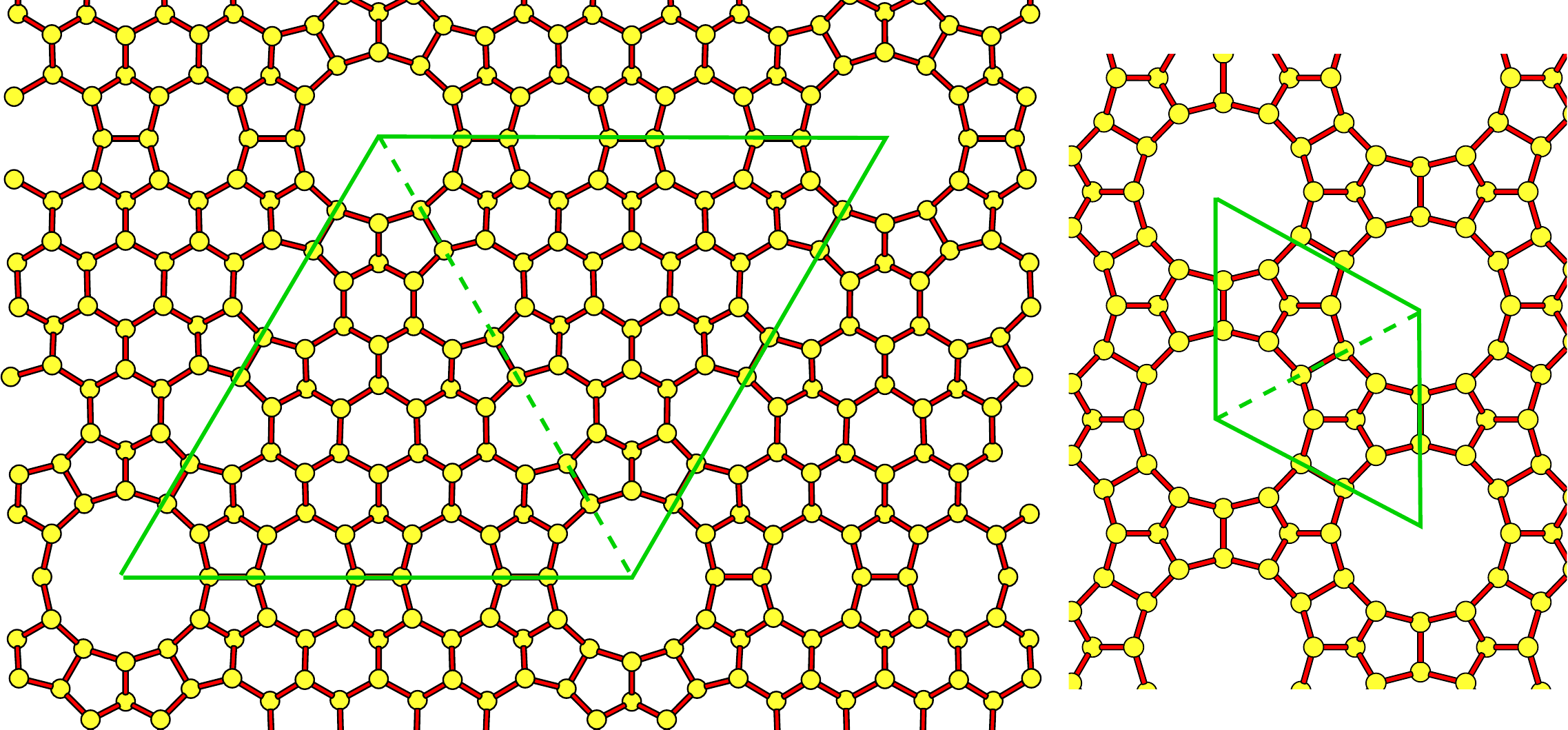}
  \caption{\label{fig:ds} {\em Left: } 7x7 DS reconstructing layer on
    the $\alpha$-Si diamond (111)-type surface. {\em Right: } 3x3 DS
    reconstructing layer, identical with a layer in clathrates.  }
\end{figure}

Below 860~C, high--temperature disordered $1\times 1$ Si(111) surface
undergoes substantial reconstruction~\cite{si77_nat}, leading
to $7\times 7$ surface super-ordering~\cite{si77_takayanagi}
(Fig.\ref{fig:ds}, left panel). The dominant motifs at the corners of
the $7\times 7$ supercell outlined green are rings of 12 pentagons
surrounding large dodecagonal ``hole'' or ``cavity''. The puckered
dodecagonal rings are accommodated by combining two perturbations of
the underlying diamond structure: ``dimerization'' along the supercell
edges and ``stacking fault'' over halves of the hexagonal supercell --
hence the ``$DS$''-model. The topmost adatomic layer completing the
known $DAS$-model of the $\alpha$-Si(111) surface termination is not
relevant for the interface studied here.

While the supercell of the $DS$ or $DAS$ model {\it must} be composed
of paired triangular areas with (faulted and unfaulted) regions
separated by triangularly shaped domain walls~\cite{si77_vanderbilt}, the
construction admits free parameter $N$ defining the triangle side
length as $(2N-1)\times (2N-1)$ supercell of the fundamental
(111)-type diamond surface periodic cell with ever sparser
(as $N$ grows) triangular lattice of the 12--fold rings. The $N=2$
case is $3\times3$ $DS$/$DAS$ reconstruction, and it reduces the
structure to dodecagonal rings : the 8--fold and 6--fold
rings present in the 7$\times$7 reconstruction are absent. While
5$\times$5 reconstruction can be readily observed for Ge deposition on the Si(111)
surface~\cite{si111ge5x5}, the 3$\times$3 reconstruction is observed rarely in
experiments due to the large compressive stress that it
suffers on the diamond(111) substrate~\cite{das_sige_zhachuk}.

\begin{figure}[t]
  \centering
  \includegraphics[width=4in,angle=0]{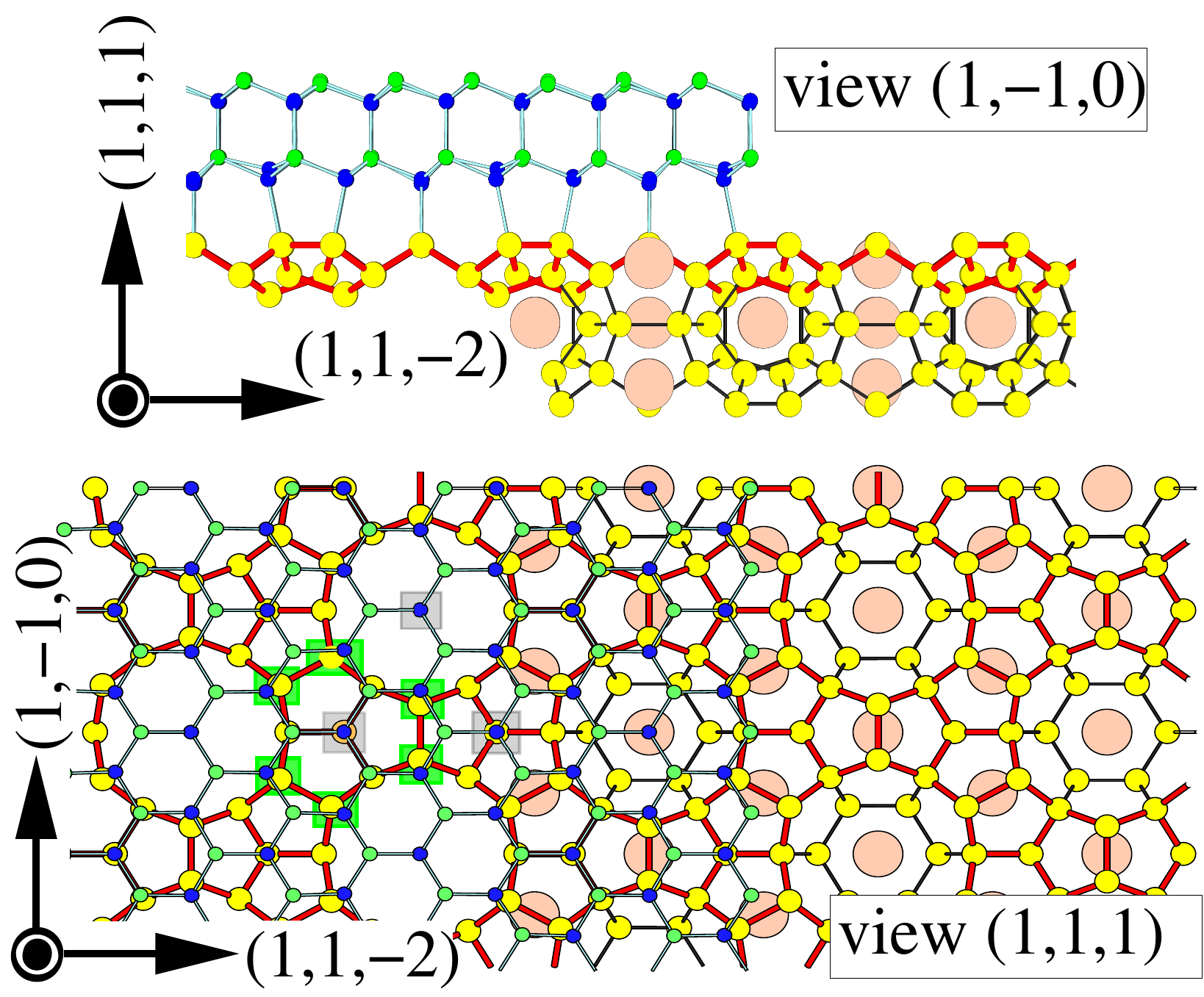}
  \caption{\label{fig:ifejs} 
    Three layers around In$_{27}$N$_{27}$/Na$_5$Si$_{34}$ interface
    viewed down the common (111) axis of the diamond and clathrate
    ({\em bottom panel}), and side view ({\em top panel}).
    Indium atoms are shown blue, nitrogen green, sodiums are large
    pink balls, and silicon is yellow. Lateral bonds within the
    clathrate layer at the interface are drawn bold and red. Green
    rectangles mark (vertical) In--Si bonds, and gray rectangles
    3--coordinated In atoms.}
\end{figure}

In systematic search for lowest--energy surface terminations of
clathrates~\cite{clay}, we observed that the best candidates are
layers normal to the 3--fold axis of the cubic clathrate ``type--II'',
or (001) of the hexagonal clathrate ``type--III'': the layer turns out to be
exactly the 3$\times$3 diamond--(111) reconstruction layer in the
right panel of the Fig.~\ref{fig:ds} -- with the caveat that embedding
in the clathrate leads to lateral misfit $>$10\%.

Figure~\ref{fig:ifejs} illustrates side and top view of the
diamond/clathrate interface on the example of quaternary
In$_{27}$N$_{27}$/Na$_5$Si$_{34}$; the red interface
layer faces the diamond side on the left, the clathrate side on the
right, while the central part shows sandwich of all three layers. On
the clathrate side, {\em all} bonds are saturated, while in the
adjacent diamond layer, six vertical bonds per hexagonal periodic cell
are saturated (green rectangles in the figure), while three are
unsaturated and protruding into hybrid interfacial cavity, highlighted
in the figure by gray rectangles.  Side view of the interface in the
top panel shows further orientational relationship: clathrate
pseudo-decagonal (1,1,-2) axis is parallel to the (1,-1,0) diamond
axis.

Lattice matching of the composite structure in Fig.~\ref{fig:ifejs}
with interface normal to (111) direction for diamond and clathrate II
structures requires that lattice parameters of the cubic diamond and
cubic clathrate type II respectively, $a_{dia}$ and
$a_{C2}$ satisfy :

\begin{equation}
  \label{eq:misfit}
  3 a_{dia} \approx a_{C2}.
\end{equation}

For the three group-IV elements Si, Ge and Sn, the generic misfit
between (guest-free) clathrate and diamond when both are constituted
by the same element is $\approx$11\%.  Clearly, by choosing
appropriate compositions for the diamond/clathrate sides of the
interface, the lattice matching can be used to favor (even guest-free)
clathrate against the competing diamond structure on one of the
interface sides. Table ~\ref{tab:misfit} shows all diamond/clathrate
compositions providing misfit $\epsilon$ less than 5\%, when combining
clathrates based on column XIV elements Si, Ge and Sn, with group IV,
III+V or II+VI element pairs for diamond side of the
heterostructure. The bottom panel shows four selected cases with
guest-atom filled clathrates. Column ``$\epsilon$'' denotes mismatch
resulting from actually measured lattice parameters, while column
``$\epsilon_{dft}$'' is DFT calculation result: the misfit minimizing
combinations are In-N/Ge (-1.2\%) and Cd-S/Sn (-0.2\%). Since
guest-free Sn clathrate--II was never prepared, we extrapolate its
hypothetical lattice parameter by calibrating DFT using $\alpha$-Sn
diamond; DFT expands its lattice by factor 1.023, hence the
``experimental'' guest-free Sn clathrate-II $a_{C2,dft}/1.023\approx
12.301$\AA. Similarly, since guest-free Si clathrate is type I not
II~\cite{chan_synthesis_2016}, we use DFT-computed ratio
$a_{C1,dft}/a_{C2,dft}$ to extrapolate $a_{C2}\approx 10.469$\AA. For
Ge, we used observed parameter $a_{C2}$=15.21\AA~from
Ref.~\cite{guloy_guest-free_2006}~and confirmed that DFT overestimates
it by 2.1\% in decent agreement with Ge-diamond case (DFT
overestimates $a_{cub}$ by 1.8\%).

Although the most conceivable experimental realization of the
interface is ``thin clathrate film on a diamond substrate'', we opted
for {\em bulk} implementation of the interface models, avoiding
possible complex reconstructions expected for both diamond and
clathrate open surfaces. In such bulk, ``sandwich'' model, both
``front'' and ``rear'' terminations of the diamond bond with a layer
of the clathrate, creating two interfaces per periodic repeat.  We
call such models {\em interface approximants}, depending on the
size/choice of the two constituting slabs. We utilize hexagonal unit
cell setting, with $a_{hex}\sim a_{C2}/\sqrt{2}\sim 3a_{dia}/\sqrt{2}$
satisfying Eq.\ref{eq:misfit}.  In this setting, the diamond portion
of the composite appears in multiples of (111)-type layer triplets,
while thinnest clathrate slab is single-cage wide fragment from cubic type II
or hexagonal type III structures. Mostly, we use the ``small''
$A_{54}/B_{34+x}$ approximant ($A$ denotes diamond part, $B$
clathrate), interlocking single triplet of (split) diamond layers with
single-cage wide layer of the clathrate. Parameter $x=5$ specifies
maximal atomic content of the clathrate ``guest'' atom eventually
occupying the cage centers. We also implemented ``medium''
$A_{54}/B_{74+x}$ and the ``large'' $A_{108}/B_{74+x}$ approximants,
containing 3 and 6 diamond layers respectively, and clathrate
represented by complete hexagonal type III unit cell motif, with
$x=12$. An example of the ``small'' approximant is in
Fig.~\ref{fig:mucenie}($a-c$), while the ``large'' approximant is shown
in Fig.~\ref{fig:mucenie}($d-e$).

\begin{table}
  \centering
  \begin{tabular}{ccc|ccc}
    & \multicolumn{2}{c}{\%} &    & \multicolumn{2}{c}{\%}\\
    dia./clath.  & $\epsilon$& $\epsilon_{dft}$&dia./clath. & $\epsilon$&$\epsilon_{dft}$\\
    \hline
    Cd-O/Si & -4.9 & -3.9 & Ge/Sn & -2.5\footnote{``experimental'' lattice parameter of empty Sn clathrate extrapolated using Sn diamond data.} & -2.7 \\
    As-B/Si & -3.2 & -2.0 &Ga-As/Sn & -2.5 & -3.0 \\
    In-N/Si & 1.6 & 2.8 &Cd-S/Sn & 0.4 & 0.1 \\
    In-N/Ge & -1.1 & -2.2 &In-As/Sn & 4.4 & 2.1 \\
    Se-Zn/Sn & -2.3 & -3.8 & & & \\
    \hline 
    Ge/NaGe  &9.8&9.9   &Ge/Cs-Sn  &-3.0&-2.7 \\
   Si/Na-Ge  &7.0 &2.2 &  As-B/Na-Si&-2.0&-2.3  \\
  \end{tabular}
  \caption{\label{tab:misfit} {\em Top panels:} Heterostructures
    diamond or zincblende/empty clathrate with lattice misfit less
    than 5\%, based on elements from group IV or group III+V or group
    II+VI for diamond. {\em Bottom panels:} selected filled-clathrate
    cases.}
\end{table}

The aim of our further DFT calculations is to prove that the
interfaces are robust against perturbations. We choose two major
tests: (1) uniaxial bond-stretching experiment along the interface
normal, and (2) isochoric heating of the heterostructure up to the
melting point. In (1), we seek maximal achievable, {\em ultimate}
stress, before the bonds start to weaken. Additionally, we are
interested how the sample cracks: does the crack always break
apart the interface constituents?

In test (2), we want to make sure that the interface model is not a
peculiar metastable state, but a robust local minimum of energy,
resistant to substantial thermal fluctuations. This is particularly
important for previously unobserved local motifs forming specifically
at the interface -- like a cage shown in Fig.\ref{fig:md}. Once the
melting occurs, we quench the sample; in our default scenario,
clathrate component of the composite will melt first; if so, the
quenched configuration should provide us with energy of competing
diamond/{\em amorphous}-state interface. Energy difference between
clathrate and quenched amorphous structure indicates likelihood of
spontaneous clathrate formation on the diamond surface.

In addition to the two tests, we compute interfacial energy $\gamma$
and separation work $W_{sep}$. Excess energy density associated with
interface area defines interfacial energy:
\begin{equation}
  \label{eq:gamma}
  \gamma = ( E_{cell} - N_cE_c - N_dE_d ) / 2S
\end{equation}
where $E_{cell}$ is energy of the ``periodic approximant'' of the
interface with $N_c+N_d$ clathrate or diamond atoms per unit cell,
$E_c$ and $E_d$ are per-atom energies of the clathrate and
diamond/zincblende bulks respectively, $S$ is interface area
and factor 2 is due to the periodic boundary. For clathrates with
guest atoms, evaluation of $E_c$ must be done at the composition of
the actual clathrate fragment constituting the interface,
hence it generally requires interpolation between two
composition-bracketing guest--atom occupancies.

By inserting at the interface a vacuum slab, we obtain {\em work of separation}, defined as
\begin{equation}
  \label{eq:wsep}
  W_{sep} = ( E_{sep} - E_{cell} ) / S \equiv  \gamma_{A} + \gamma_{B} - \gamma_{ifc}
\end{equation}
where $E_{cell}$ is per-cell energy of the (relaxed) interface, 
$E_{sep}$ is its energy upon inserting the vacuum slab, and
$\gamma_{A,B}$ are the two related surface energies.

An important caveat behind our test calculations is that the interface
approximants involving ``binary diamond--type'' zincblende structures with $A^1A^2$
composition are {\em polar}, with two inequivalent surfaces/interfaces
$A^1$--$B$ or $A^2$--$B$ (clathrate host based on element $B$). While
computing {\em average} quantities like $\gamma$ following
Eq.~\ref{eq:gamma} is straightforward, decoupled quantities
$\gamma_{A^1B}$ and $\gamma_{A^2B}$ require nontrivial effort --
hydrogen-passivation or ``tetrahedral cluster''
approach~\cite{polar_surface}. Similarly, it is easy to compute
decoupled quantity like $W_{sep-A^1B}$ or $W_{sep-A^2B}$ following
Eq.~\ref{eq:wsep}, by inserting vacuum slab at a given interface;
however, due to the polar constraint the surface vs binding
contributions into $W_{sep}$ remain ambiguous. The tensile straining
experiment is also affected by the polar ambiguity, but the computed
ultimate stresses are always due to the {\em weaker} interface.  In
this paper, we do not elaborate on resolving the polar ambiguity,
since the fundamental properties of the interfaces -- lattice
mismatch, and robustness of the interface stability -- remain
unaffected.

\section{Results and Discussion}
\label{sec:results}

Summary of the data resulting from diamond/clathrate interface testing
simulations are gathered in Table~\ref{tab:data}. All rows except the
case $M$-Ge/CsSn refer to ``small'' interface approximants
$A_{54}/B_{34+x}$ (diamond--structure atoms $A$, clathrate $B$, per
unit cell), and optional $x=5$ content of the guest atom in the
clathrate; in case of zinc-blende, the formula is
$A^1_{27}A^2_{27}/B_{34+x}$.  Row $M$-Ge/CsSn refers to
``medium''-size approximant $A_{54}B_{72+x}$, $x=12$.

\begin{figure}[t]
  \centering
  \includegraphics[width=4in]{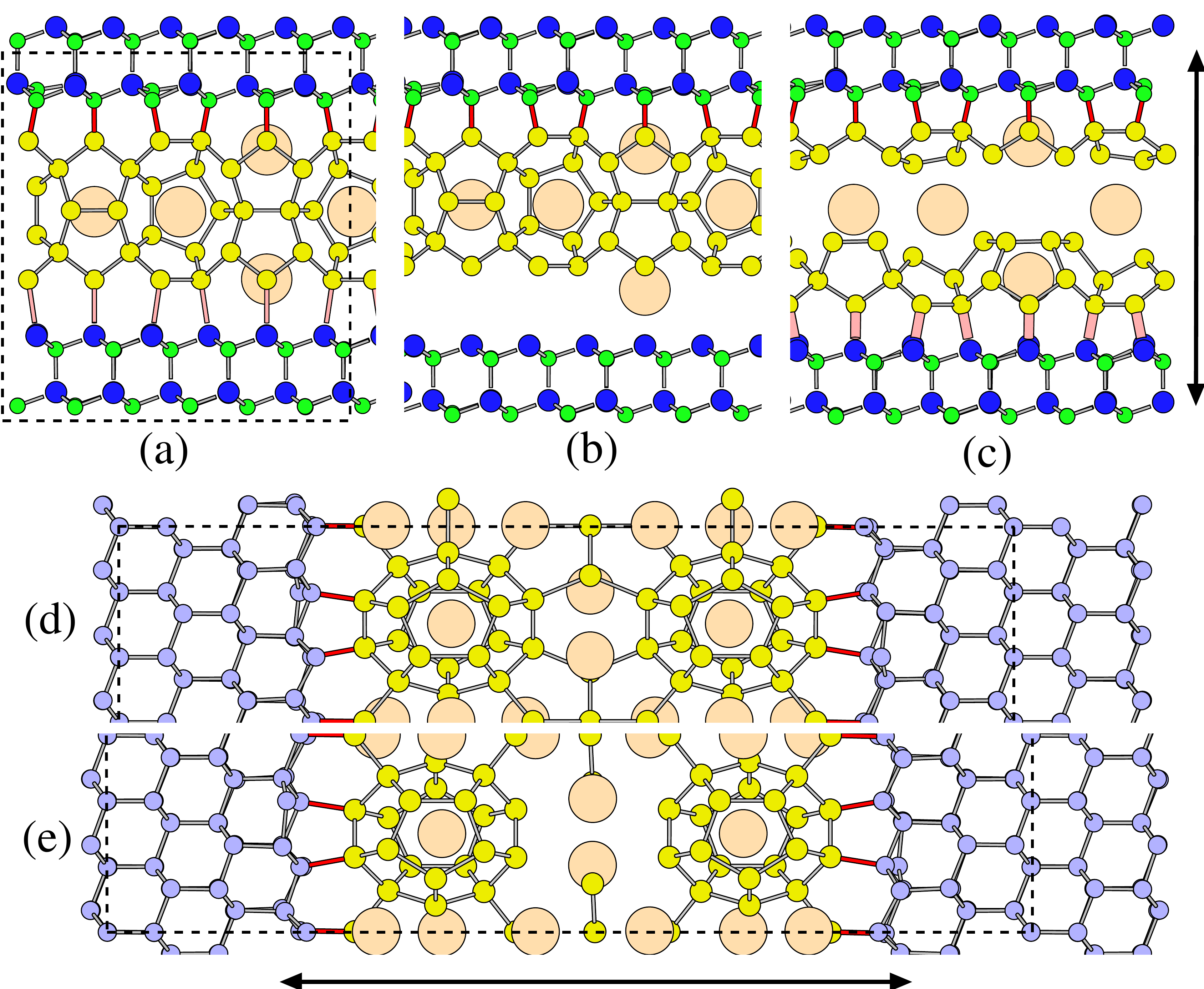}      
  \caption{\label{fig:mucenie} {\em Top panels (a-c)}: AsB/NaSi
    interface at its ultimate strength (strain $\epsilon_u$=10\%) {\em
      (a)}, after failure by breaking pink As--Si bonds
    ($\epsilon_u$=12\%) {\em (b)}, and after failure by breaking Si-Si
    intra-clathrate bonds when (thick pink) As-Si bonds had fixed
    vertical height ($\epsilon_u$=14\%) {\em (c)}. {\em Bottom panels
      (d-e)}: Ge/CsSn large--approximant interface at ultimate
    strength ($\epsilon_u$=13\%) {\em (d)}, and after failure by
    intra-clathrate crack ($\epsilon_u$=16\%) {\em (e)}. Atoms color
    code: As -- blue; B -- green; Na and Cs : light-orange ; Si, Sn --
    yellow. Arrows indicate axial straining direction.  }
\end{figure}

First of all, the recorded ultimate stresses $\sigma_u$ suggest that
the interface components are bonded: smallest ultimate stress 54~kB
occurs for ($M$-Ge/CsSn), but in that case the last column ``crack''
indicates intra-clathrate Sn--Sn bond breaking. In one case (AsB/NaSi)
illustrated in Fig.\ref{fig:mucenie}{\em a-c}, after the crack first
occured at As--Si interface ($b$), we repeated the stretching with the
six As-Si bond pairs fixed as indicated by thick pink sticks in
($c$). Consequently, we observed intra-clathrate crack at
$\sigma_u\sim 152$~kB, suggesting that the B-Si interface bonds (red in
Fig.\ref{fig:mucenie}$c$) are stronger than the Si-Si bonding of the
most compact dodecahedral cages. Bottom panel of the same figure shows ``large''
interface approximant Ge/CsSn with 6 $\alpha$-Ge (111)-diamond layers,
and clathrate represented by entire hexagonal type III cell
content: in ($d$), the interface is shown at ultimate stress
$\sigma_u\sim 54$~kB, and panel ($e$) shows intra-clathrate crack
developing upon further increase of the strain. Exactly the same
picture was observed for ``medium''-sized approximant $M$-Ge/CsSn
reported in Tab.~\ref{tab:data} with identical clathrate component,
but only 3 diamond--type layers. Note that in the Figure, the
stretching direction is {\em vertical} for top-panel ($a$-$c$), and
{\em horizontal} in ($d$-$e$).

To further illustrate the interface bonding strength,
Table~\ref{tab:ustress} gathers reference ultimate stresses for
selected diamond-type structures, and guest-free clathrate type II. In
case of silicon clathrate, the weaker-interface (In-Si, As-Si)
$\sigma_u$ is about 2/3 of the intra-clathrate Si value, and about
half of the intra-diamond (In-N, As-B) values. However, in case of Ge
and Sn based clathrates, the intra-clathrate $\sigma_u$ values are
comparable to the interfacial Ge-Si or Ge-In bond strengths.

AIMD heating simulations were performed to ($i$) test robustness of
the interface by thermal perturbation; and ($ii$) by
quenching the melted clathrate substructure (assuming the diamond-type
slab remained intact), assess energy difference $\Delta
E_a=E_{amorph}-E_{clath}$ per clathrate atom -- we assume
that amorphous structure is the only potential clathrate competitor,
neutral vs lattice-matching advantage of the clathrate. Note that Ge
or Si {\em amorphous} phases were extensively studied~\cite{asi,age},
while mentions of the amorphous Sn are nearly absent in the
literature: this is of interest whenever we demand guest-free
clathrate.

We find that the interface melting generally occurs at temperatures
comparable with melting temperatures of its constituents.  Ge/Sn
composition partially melted above $T\sim$500K, comparable to
$\beta$-tin melting temperature, Ge--diamond substructure remained
stable. Quenching produced interface with amorphized Sn substructure with {\em the same}
energy as the initial interface,
indicating little chance to prepare guest-free Sn clathrate on the
Ge(111) surface. Indeed, experimentally Sn on Ge(111) is amorphous
above 1.6~ML coverage at least up to 2~ML~\cite{sn_on_ge111}. On the other hand, chances
to prepare guest-free Sn clathrate on lattice-matched CdS or SeZn
diamond substrates are good: not only the Sn clathrate remains
mechanically stable up to 950-1000K, but the quenched/relaxed
amorphous layer is substantially unstable to the clathrate. In case of
Cs-filled clathrate, {\em both} substructures melt simultaneously,
well {\em above} the reported decomposition temperature $\sim$850K of
the Cs$_8$Sn$_{44}$ clathrate~\cite{clath-cssn}.

For the experimentally prepared composition Ge/NaGe~\cite{kume_gena},
both interface components rather surprisingly melted simultaneously
at 900K after 20 ps annealing. Experimentally, the clathrate is only
metastable, and stable Ge$_4$Na melts around
1200K~\cite{pdiag-gena}. It appears that Ge-Na clathrate formation on
Ge(111) could be also due to lattice matching - despite the 11\%
mismatch. Guest-free Ge clathrate on InN surface is an example of well
matched, stable interface, and we predict that guest-free clathrate
type II should spontaneously form on InN(111). Interestingly, cubic
diamond-type zincblende InN phase can be prepared on
Yttrium-stabilized zirconia (YSZ)~\cite{inn-on-yzr} with excellent
lattice matching, hence Ge clathrate should also enjoy lattice
matching advantage on YSZ.  (Clathrate formation on sapphire substrate
proved viable despite the apparent $>20\%$
mismatch~\cite{gena-on-sapphire}).

We also studied Si-host clathrate interfaced to two alternative cubic
zincblende structures: AsB and InN. The zincblende(AsB) remained
stable upon clathrate melting at 1700K, and the quenched sample
displayed amorphous Na--Si structure with positive $\Delta E_a$ of 172
meV per (Na+Si) atom respectively, suggesting feasibility of Na-Si
clathrate formation on (111)-AsB surface.  The InN/NaSi interface
endured up to high $T=$1550~K annealing; melting occured after
$\sim$15 ps at 1600 K by diffusive penetration of Nitrogen into the
clathrate, and strong reaction with Si, presumably an early stage of
 nucleating Si$_3$N$_4$.

\begin{table}
  \centering
  \begin{threeparttable}
  \begin{tabular}{ccccccc}
    interface & $W_{sep}$ &$\gamma$ & $T_m$ & $\Delta E_a$ & $\sigma_u$ & crack \\
dia./clath.   &\multicolumn{2}{r}{meV/\AA$^2$}     & K      &  meV/at.   & kB & type \\
\hline
Ge/Sn           & 64       & 24 & 550 & 0     & 64 & Ge--Sn  \\  
CdS/Sn      & 67, 80   & 16 & 1000    & 186         & 65 & Cd--Sn \\  
SeZn/Sn      & 78, 75    & 20 & 950   & 176         & 62 & Se--Sn \\   
Ge/CsSn         & 63       & 14 &1000 & 187\tnote{a} & 60 & Ge--Sn \\ 
$M$-Ge/CsSn\tnote{b}& 65       & 18 &1050 & 171\tnote{a} & 54 & Sn--Sn \\
Ge/NaGe     & 72       & 23 & 900  & 154\tnote{a}& 79  & Ge--Ge \\
InN/Ge      & 96, 162  & 27 & 1250 & 217         & 108 & In--Ge \\ 
Si/NaGe     & 78       & 23 & 1150 & 67   & 104 & Si--Ge \\ 
AsB/NaSi    & 111, 336 & 28 &1700  & 172 & 131 & As--Si \\ 
InN/NaSi    & 98, 208  & 5  & 1650 &-307\tnote{c}    & 108 & In--Si \\
\hline
  \end{tabular}
  \begin{tablenotes}
  \item[a]{both interface components melted simultaneously}
  \item[b]{``medium''--size approximant Ge$_{54}$/Cs$_{12}$Sn$_{74}$}
  \item[c]{Nitrogen diffused into clathrate}
  \end{tablenotes}
  \caption{\label{tab:data} Separation work $W_{sep}$, average
    interfacial energy $\gamma$, melting temperature $T_m$ and $\Delta
    E_a$ -- energy per atom of the melted/quenched (amorphized) interface sandwich
    relative to the initial diamond/clathrate reference state. The last two
      columns are ultimate axial strength $\sigma_u$ and bond--type broken by crack.}
  \end{threeparttable}
\end{table}

\begin{table}[t]
  \begin{tabular}{cccccccc}
        & Si    & Ge    & Sn   & Cd-S & Se-Zn & In-N  & As-B \\
diamond & 189 & 121 & 73 & 88 & 89  & 232 & 260 \\    
\hline
clath.II  & 170 & 101 & 59 & & & & \\
  \end{tabular}
  \caption{\label{tab:ustress}
    Ultimate axial stress in kB for selected diamond/zincblende structures, and clathrates (bottom row).
  }
\end{table}

\begin{figure}[t]
  \centering
  \includegraphics[width=3.1in]{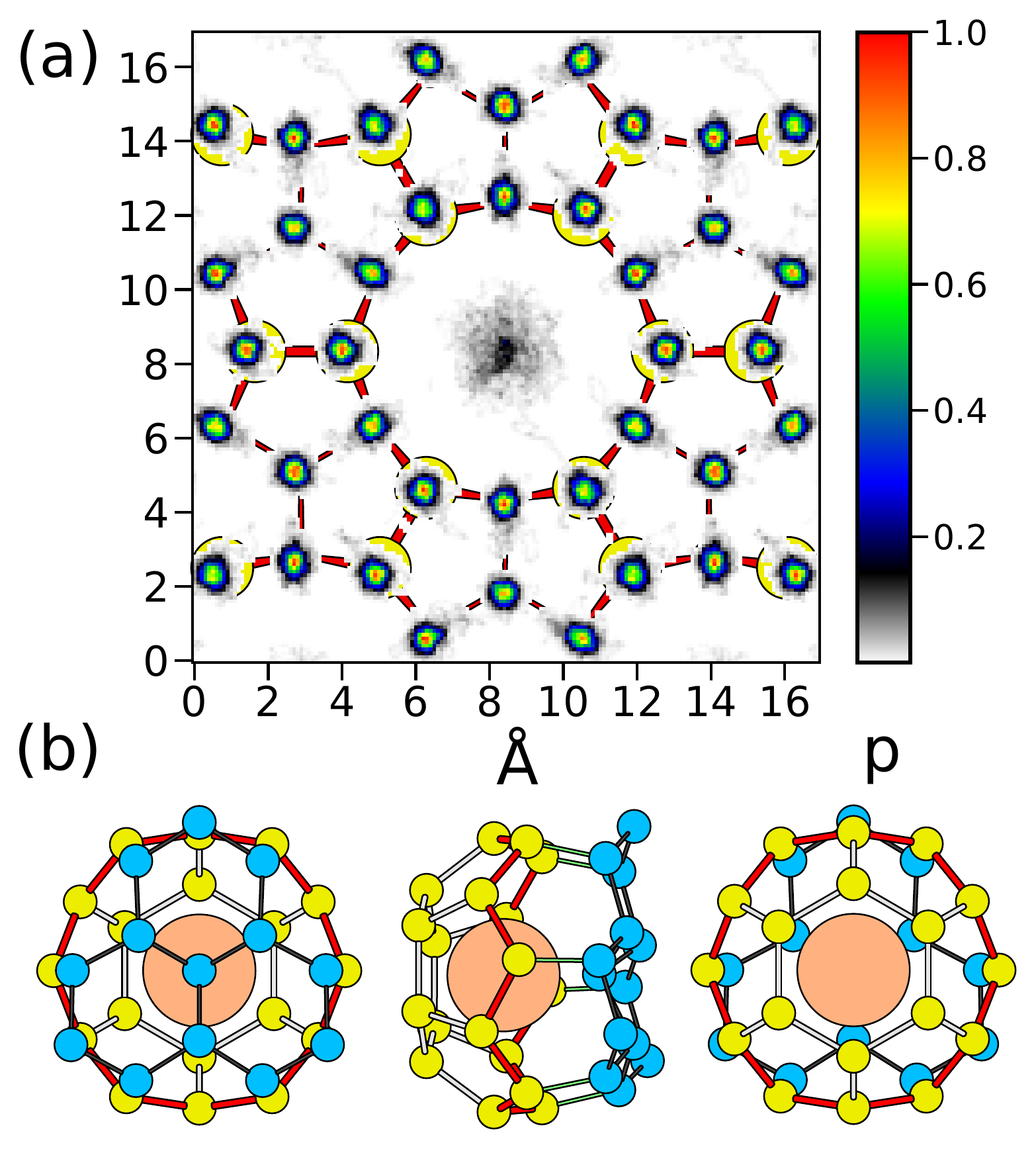}
  \caption{\label{fig:md} {\em Top panel (a):} Probability density $p$
    of atomic occupancy within a 4\AA--thick slice through $Si/GeNa$
    interface, obtained from $T$=1100K AIMD simulation. The
    distribution overlays ball-stick model, referring to puckered
    12--fold rings in red-stick layers of Fig.~\ref{fig:ifejs} or
    Fig.~\ref{fig:ds} (right panel). Bottom panel~$(b$): front, side and
    rear views of the ``hybrid cage'' at the interface. Red sticks
    show 12-fold ring from top panel $(a)$. }
\end{figure}
The interface-specific structural feature is hybrid cage with cavity
on the clathrate side, covered by ``flat'' atomistic configuration
belonging to the first complete diamond layer next to the interface
(Fig.~\ref{fig:md}$(b)$). The diamond-side cage boundary (cyan atoms)
is centred by an atom with one unsaturated bond (marked by gray
squares in the visualization of Fig.~\ref{fig:ifejs}), protruding
toward cage interior toward the guest atom. Fig.~\ref{fig:md}$(a)$
illustrates stability of this object during high--temperature
annealing: in the top panel, occupancy distribution in a 4~\AA--thick
slice through the center of the cage is superimposed on the ball-stick
model. The distribution obtained by accumulating pixelled occupancy
statistics of the $\sim$50ps $T$=1100K AIMD trajectory for the example
of Si/GeNa interface composition, was normalized to maximal observed
occupancy count. The guest--atom positional distribution is broad, but
centred; the clathrate host positions are sharply defined, although
the occasional dark-gray features nearby indicate closeness of the
melting transformation (at 1150 K). Thus the hybrid cage appears to be
mechanically very stable part of the interface.

Columns 2--3 of the Table~\ref{tab:data} show separation work
$W_{sep}$ and average interfacial energies $\gamma$. For zincblende
(``binary diamond-type''), we quote two values of $W_{sep}$; the first
of them is always for the weaker-bonded interface variant; normally,
these are bond types broken in cracking event during stretching
simulation, and reported in the last column. The by far largest
$W_{sep}$ is observed for B-Si bonded interface of AsB/NaSi -- this
looks remarkable since there is no Si-rich B--Si compound.  $W_{sep}$
for N--Si on the other hand is also high, but in this case (DFT
computed) mixing enthalpy of Si$_3$N$_4$ compound is substantial, -1.1
eV/atom. In order to elucidate the values of $W_{sep}$, we computed
the same quantity also for guest-free Si, Ge and Sn clathrate type II,
by introducing vacuum slab normal to (111) direction, and obtained
124, 98 and 64 meV/A$^2$ respectively. Thus, the cost of inserting
vacuum slab at an interface is comparable to intra-clathrate
$W_{sep}$. Another comparison can be made with recently reported
$W_{sep}\sim$60~meV/\AA$^{\rm 2}$ for Sn decagonal clathrate film on
metallic decagonal AlCoNi quasicrystal~\cite{sn_on_alconi} -- although
in the present case we do not disambiguate contributions of the
bond--breaking vs surface energies. On the other hand, these values
are a fraction of $W_{sep}$ in metallic systems: $\sim$140 meV/\AA$^{\rm 2}$
for Ag/Fe interface~\cite{ifc-agfe}, or 320 meV/\AA$^2$ for AlNi/Cr
interface~\cite{ifc-alnicr}.

\begin{table}
  \begin{tabular}{cccccc}
 GaP/Si    & SZn/Si    & AsGa/Ge  & SeZn/Ge & InSb/Sn & CdTe/Sn \\
    7.5 & 33.2 & 6.5 & 74.3 & 16.3 & 13.0 \\    
  \end{tabular}
  \caption{\label{tab:ogammas}
    Average (111)--type zincblende/diamond interfacial energies in meV/\AA$^{\rm 2}$
  }
\end{table}

The average interfacial energy $\gamma$ (column 3 in
Tab.~\ref{tab:data}) ranges between 5--28 meV/\AA$^{\rm 2}$, the
extremes are due to InN/NaSi and AsB/NaSi cases respectively. The very
low $\gamma$ for InN/NaSi can be due to large formation enthalpy of
silicon nitride noted above: N-Si bonds are highly favorable while
formation of the Si$_3$N$_4$ is kinetically suppressed. As noted in
Sec.~\ref{sec:ifejs}, the interfaces with zincblende are polar, hence
the computed $\gamma$'s are just averages; nevertheless we can still
meaningfully compare them with similarly lattice-matched
(111)-zincblende/(111)-diamond interfaces not exhibiting the abrupt
structural change. Table~\ref{tab:ogammas} summarizes $\gamma$'s for
such interfaces with nearly-zero lattice mismatch. Ranges of $\gamma$
values are about the same as in case of zincblende/clathrate
interfaces, with a single exception of the SeZn/Ge case with
surprisingly high interfacial energy, confirming that heterostructural
interfaces with clathrate are indeed energetically competitive. To
contextualize these $sp^3$--system interfacial energies, the
aforementioned metallic system's $\gamma$'s are 23 for Ag/Fe and 37
meV/\AA$^{\rm 2}$ for AlNi/Cr. For chemically different carbid/Fe
interfaces, the $\gamma$'s can be as small as 16 for Fe/TiC or even 10
meV/\AA$^{\rm 2}$ for Fe/ZrO~\cite{ifc-fec}, and for Lithium-based
interfaces Li$_2$O or Li$_2$S $\gamma$=26 and 11 meV/\AA$^{\rm 2}$
respectively~\cite{ifc-li}. Overall, all of these values are just in
the range of the interfacial energies of our model clathrate
interfaces.

\section{Summary and Conclusion}
\label{sec:summary}

The generic lattice mismatch $3a_{dia}/a_{C2}\sim 1.1$ between
(111)-oriented terminations of diamond and clathrate type II
structures can be turned into selective energetic advantage for the
matching clathrate structure, upon appropriate choice of the interface
side compositions. Such lattice matched interfaces give rise to a
hybrid pseudocages that can fit a large guest atom. The interfaces are
very stable against thermal fluctuations and tightly $sp^3$--bonded,
as is clear from the computational axial tensile stress experiments.

The geometrical principle of the proposed diamond/clathrate interface
-- matching between (111)-diamond honeycomb lattice and the ``3x3
DS''--like layer that is part of some clathrates -- can be generalized
to alternative structural families.
An example of such (111)-type
termination is cubic $B2$ structure satisfying matching condition
$3\sqrt{2} a_{B2} \sim a_{C2}$ where $a_{B2}$ is lattice parameter of
the $B2$ structure. A concrete example could be Al--terminated
AlRh($B2$) with Sn(clathrate-II). Oxides are another class of
materials with a potential to serve as a substrate for epitaxial
clathrate films~\cite{gena-on-sapphire}, despite apparent lack of
lattice-matching between sapphire and GeNa clathrate type II. On the
other hand, cubic-InN matches near-perfectly with Ge empty clathrate
(see Tab.~\ref{tab:misfit}), but also with zirconia~\cite{inn-on-yzr},
suggesting nice lattice-matching between cubic-ZrO$_2$ and
Ge(clathrate-II). In the latter case, the unsaturated Ge-clathrate
bonds at the interface might not easily interact with electronically
balanced ZrO$_2$ sheets, but the standard adjustment using
Zr$\rightarrow$Y substitution could help revealing the
lattice-matching potential between yttrium-stabilized zirconia (YSZ)
and Ge clathrate.

\section{ Acknowledgements}
We acknowledge support from grants VEGA 2/0144/21, APVV-19-0369 and
APVV-20-0124.  M.M. also acknowledges support of the CEMEA project:
Building-up Centre for advanced materials application of the Slovak
Academy of Sciences, ITMS project code 313021T081 supported by the
Integrated Infrastructure Operational Programme funded by the ERDF
(European Regional Development Fund, EU).  Calculations were performed
in the Computing Center of the Slovak Academy of Sciences using the
supercomputing infrastructure acquired under Projects ITMS 26230120002
and 26210120002.

\bibliographystyle{elsarticle-num} 
\bibliography{iface}

@article{kume_gena,
	title = {A thin film of a type {II} {Ge} clathrate epitaxially grown on a {Ge} substrate},
	volume = {18},
	number = {30},
	journal = {CrystEngComm},
	author = {Kume, Tetsuji and Ban, Takayuki and Ohashi, Fumitaka and Jha, Himanshu S. and Sugiyama, Tomoya and Ogura, Takuya and Sasaki, Shigeo and Nonomura, Shuichi},
	year = {2016},
	pages = {5630--5638},
}

@article{iceclath,
	title = {Can clathrates heterogeneously nucleate ice?},
	volume = {151},
	number = {11},
	journal = {The Journal of Chemical Physics},
	author = {Factorovich, Matías H. and Naullage, Pavithra M. and Molinero, Valeria},
	year = {2019},
	pages = {114707},
}

@article{patchypart,
	title = {Assembly of clathrates from tetrahedral patchy colloids with narrow patches},
	volume = {151},
	number = {9},
	journal = {The Journal of Chemical Physics},
	author = {Noya, Eva G. and Zubieta, Itziar and Pine, David J. and Sciortino, Francesco},
	year = {2019},
	pages = {094502},
}

@article{si77_nat,
	author = {Kitamura, Si. and Sato, T. and Iwatsuki, M.},
	title = {Observation of surface reconstruction on silicon above 800 C using the STM},
	journal = {Nature},
        volume={351}, 
	pages = {215--217}, 
	year = {1991},
}

@article{si77_takayanagi,
	title = {Structural analysis of {Si}(111) by {UHV} transmission electron diffraction and microscopy},
	volume = {3},
	number = {3},
	journal = {Journal of Vacuum Science \& Technology A: Vacuum, Surfaces, and Films},
	author = {Takayanagi, K. and Tanishiro, Y. and Takahashi, M. and Takahashi, S.},
	month = jun,
	year = {1998},
	pages = {1502},
}

@article{si77_vanderbilt,
	title = {Model for the energetics of {Si} and {Ge} (111) surfaces},
	volume = {36},
	number = {11},
	journal = {Physical Review B},
	author = {Vanderbilt, David},
	month = oct,
	year = {1987},
	pages = {6209},
}

@article{si111ge5x5,
  title = {Surface electronic structure of Si(111)7\ifmmode\times\else\texttimes\fi{}7-Ge and Si(111)5\ifmmode\times\else\texttimes\fi{}5-Ge studied with photoemission and inverse photoemission},
  author = {Martensson, Per and Ni, Wei-Xin and Hansson, G\"oran V. and Nicholls, J. Michael and Reihl, Bruno},
  journal = {Phys. Rev. B},
  volume = {36},
  issue = {11},
  pages = {5974--5981},
  year = {1987},
  month = {Oct},
}

@article{das_sige_zhachuk,
	title = {Strain-induced structure transformations on {Si}(111) and {Ge}(111) surfaces: a combined density-functional and scannning tunnneling microscopy report},
	volume = {138},
	shorttitle = {Strain-induced structure transformations on {Si}(111) and {Ge}(111) surfaces},
	number = {22},
	journal = {The Journal of Chemical Physics},
	author = {Zhachuk, R. and Teys, S. and Coutinho, J.},
	month = jun,
	year = {2013},
	pages = {224702},
}

@unpublished{clay,
	title={Stable thin clathrate layers},
	author={Pospíšilová, E. and Mihalkovič, M.},
	year={2022},
	note={unpublished}
}

@ARTICLE{PAW,
  author =       {G. Kresse and D. Joubert},
  title =        {From ultrasoft pseudopotentials to the projector augmented-wave method},
  year =         {1999},
  journal =      {Phys. Rev. B},
  volume =       {59},
  pages =        {1758-75}
}

@Article{PW91,
  author =       {J. P. Perdew and J. A. Chevary and S. H. Vosko and K. A. Jackson and M. R. Pederson and D. J. Singh and C. Fiolhais},
  title =        {Atoms, molecules, solids, and surfaces: Applications of the generalized gradient approximation for exchange and correlation},
  journal =      {Phys. Rev. B},
  year =         {1992},
  OPTkey =       {},
  volume =       {46},
  OPTnumber =    {},
  pages =        {6671-87},
  OPTmonth =     {},
  OPTnote =      {},
  OPTannote =    {}
}

@ARTICLE{Kresse96,
  author =       {G. Kresse and J. Furthmuller},
  title =        {Efficient iterative schemes for ab initio total-energy calculations using a plane-wave basis set},
  year =         {1996},
  journal =      {Phys. Rev. B},
  volume =       {54},
  pages =        {11169-86}
}

@article{chan_synthesis_2016,
	title = {Synthesis and {Characterization} of {Empty} {Silicon} {Clathrates} for {Anode} {Applications} in {Li}-ion {Batteries}},
	volume = {1},
	issn = {2059-8521},
	url = {http://link.springer.com/10.1557/adv.2016.434},
	doi = {10.1557/adv.2016.434},
	language = {en},
	number = {45},
	urldate = {2022-05-24},
	journal = {MRS Advances},
	author = {Chan, Kwai S. and Miller, Michael A. and Ellis-Terrell, Carol and Chan, Candace K.},
	month = sep,
	year = {2016},
	pages = {3043--3048},
}

@article{guloy_guest-free_2006,
	title = {A guest-free germanium clathrate},
	volume = {443},
	issn = {0028-0836, 1476-4687},
	url = {http://www.nature.com/articles/nature05145},
	doi = {10.1038/nature05145},
	language = {en},
	number = {7109},
	urldate = {2021-11-04},
	journal = {Nature},
	author = {Guloy, Arnold M. and Ramlau, Reiner and Tang, Zhongjia and Schnelle, Walter and Baitinger, Michael and Grin, Yuri},
	month = sep,
	year = {2006},
	pages = {320--323},
}

@article{polar_surface,
	title = {Pseudo-{Hydrogen} {Passivation}: {A} {Novel} {Way} to {Calculate} {Absolute} {Surface} {Energy} of {Zinc} {Blende} (111)/(-1-1-1) {Surface}},
	volume = {6},
	issn = {2045-2322},
	shorttitle = {Pseudo-{Hydrogen} {Passivation}},
	url = {http://www.nature.com/articles/srep20055},
	doi = {10.1038/srep20055},
	language = {en},
	number = {1},
	urldate = {2022-11-26},
	journal = {Scientific Reports},
	author = {Zhang, Yiou and Zhang, Jingzhao and Tse, Kinfai and Wong, Lun and Chan, Chunkai and Deng, Bei and Zhu, Junyi},
	month = apr,
	year = {2016},
	pages = {20055},
}

@article{clath-cssn,
	title = {Order-{Disorder} {Phase} {Transition} in {Type}-{I} {Clathrate} {Cs8Sn44}},
	volume = {2007},
	issn = {14341948, 10990682},
	url = {https://onlinelibrary.wiley.com/doi/10.1002/ejic.200700644},
	doi = {10.1002/ejic.200700644},
	language = {en},
	number = {26},
	urldate = {2021-11-04},
	journal = {European Journal of Inorganic Chemistry},
	author = {Kaltzoglou, Andreas and Hoffmann, Stephan D. and Fässler, Thomas F.},
	month = sep,
	year = {2007},
	pages = {4162--4167},
}

@article{pdiag-gena,
	title = {Thermodynamic description of the {Ge}--{Na} and {Ge}--{K} systems using the {CALPHAD} approach supported by first-principles calculations},
	volume = {37},
	issn = {03645916},
	url = {https://linkinghub.elsevier.com/retrieve/pii/S0364591612000065},
	doi = {10.1016/j.calphad.2012.01.004},
	language = {en},
	urldate = {2022-11-30},
	journal = {Calphad},
	author = {Wang, Yaru and Wang, Peisheng and Zhao, Dongdong and Hu, Biao and Du, Yong and Xu, Honghui and Chang, Keke},
	month = jun,
	year = {2012},
	pages = {72--76}
}

@article{sn_on_ge111,
	author = {M. Göthelid and T.M. Grehk and M. Hammar and U.O. Karlsson and S.A. Flodström},
	title = {Adsorption of tin on the {Ge}(111)-c(2x8) surface studied with scanning tunneling microscopy and photoelectron spectroscopy},
	volume = {328},
	issn = {0039-6028},
	url = {https://www.sciencedirect.com/science/article/abs/pii/0039602895000267},
	doi = {10.1016/0039-6028(95)00026-7},
	language = {en},
	number = {1-2},
	urldate = {2022-05-24},
	journal = {Surface Science},
	month = apr,
	year = {1995},
	note = {Publisher: North-Holland},
	pages = {80--94}
}

@article{inn-on-yzr,
	title = {Field-effect transistors based on cubic indium nitride},
	volume = {4},
	issn = {2045-2322},
	url = {http://www.nature.com/articles/srep03951},
	doi = {10.1038/srep03951},
	language = {en},
	number = {1},
	urldate = {2022-11-30},
	journal = {Scientific Reports},
	author = {Oseki, Masaaki and Okubo, Kana and Kobayashi, Atsushi and Ohta, Jitsuo and Fujioka, Hiroshi},
	month = may,
	year = {2015},
	pages = {3951},
}

@article{gena-on-sapphire,
	title = {A fabrication method for type-{II} {Ge} clathrate film by annealing of {Ge} film covered with {Na} layer},
	volume = {734},
	issn = {00406090},
	url = {https://linkinghub.elsevier.com/retrieve/pii/S0040609021003424},
	doi = {10.1016/j.tsf.2021.138859},
	language = {en},
	urldate = {2022-05-24},
	journal = {Thin Solid Films},
	author = {Kumar, Rahul and Hazama, Yuta and Ohashi, Fumitaka and Jha, Himanshu S. and Kume, Tetsuji},
	month = sep,
	year = {2021},
	pages = {138859},
}

@article{asi,
  title = {Evidence of Voids Within the As-Deposited Structure of Glassy Silicon},
  author = {Moss, S. C. and Graczyk, J. F.},
  journal = {Phys. Rev. Lett.},
  volume = {23},
  issue = {20},
  pages = {1167--1171},
  numpages = {0},
  year = {1969},
  month = {Nov},
  publisher = {American Physical Society},
  doi = {10.1103/PhysRevLett.23.1167},
  url = {https://link.aps.org/doi/10.1103/PhysRevLett.23.1167}
}

@article{age,
author = { W.   Paul  and  G.A.N.   Connell  and  R.J.   Temkin },
title = {Amorphous germanium I. A model for the structural and optical properties},
journal = {Advances in Physics},
volume = {22},
number = {5},
pages = {531-580},
year  = {1973},
publisher = {Taylor \& Francis},
doi = {10.1080/00018737300101339}
}

@article{sn_on_alconi,
  title = {Decagonal Sn clathrate on $d$-Al-Ni-Co},
  author = {Singh, Vipin Kumar and Posp\'{\i}\ifmmode \check{s}\else \v{s}\fi{}ilov\'a, Eva and Mihalkovi\ifmmode \check{c}\else \v{c}\fi{}, Marek and Kraj\ifmmode \check{c}\else \v{c}\fi{}\'{\i}, Marian and Bhakuni, Pramod and Sarkar, Shuvam and Pussi, Katariina and Schlagel, D. L. and Lograsso, T. A. and Canfield, Paul C. and Roy Barman, Sudipta},
  journal = {Phys. Rev. B},
  volume = {107},
  issue = {4},
  pages = {045410},
  numpages = {21},
  year = {2023},
  month = {Jan},
  publisher = {American Physical Society},
  doi = {10.1103/PhysRevB.107.045410},
  url = {https://link.aps.org/doi/10.1103/PhysRevB.107.045410}
}

@ARTICLE{martinez2013,
  author={Martinez, Aaron D. and Krishna, Lakshmi and Baranowski, Lauryn L. and Lusk, Mark T. and Toberer, Eric S. and Tamboli, Adele C.},
  journal={IEEE Journal of Photovoltaics},
  title={Synthesis of Group IV Clathrates for Photovoltaics},
  year={2013},
  volume={3},
  number={4},
  pages={1305-1310},
  doi={10.1109/JPHOTOV.2013.2276478}
}

@article{ifc-li,
  title={Modeling interfaces between solids: application to Li battery materials},
  author={Lepley, ND and Holzwarth, NAW},
  journal={Physical Review B},
  volume={92},
  number={21},
  pages={214201},
  year={2015},
  publisher={APS}
}

@article{ifc-fec,
  title={Energetics for interfaces between group IV transition metal carbides and bcc iron},
  author={Jung, Woo-Sang and Lee, Seung-Cheol and Chung, Soon-Hyo},
  journal={ISIJ international},
  volume={48},
  number={9},
  pages={1280--1284},
  year={2008},
  publisher={The Iron and Steel Institute of Japan}
}

@article{ifc-alnicr,
  title = {$\mathrm{Ni}\mathrm{Al}(110)/\mathrm{Cr}(110)$ interface: A density functional theory study},
  author = {Liu, W. and Li, J. C. and Zheng, W. T. and Jiang, Q.},
  journal = {Phys. Rev. B},
  volume = {73},
  issue = {20},
  pages = {205421},
  year = {2006},
  month = {May},
  publisher = {American Physical Society},
  doi = {10.1103/PhysRevB.73.205421},
  url = {https://link.aps.org/doi/10.1103/PhysRevB.73.205421}
}

@article{ifc-agfe,
  title = {First-principles study of fcc-Ag/bcc-Fe interfaces},
  author = {Lu, Song and Hu, Qing-Miao and Punkkinen, Marko P. J. and Johansson, B\"orje and Vitos, Levente},
  journal = {Phys. Rev. B},
  volume = {87},
  issue = {22},
  pages = {224104},
  numpages = {11},
  year = {2013},
  month = {Jun},
  publisher = {American Physical Society},
  doi = {10.1103/PhysRevB.87.224104},
  url = {https://link.aps.org/doi/10.1103/PhysRevB.87.224104}
}

\end{document}